\documentclass[twocolumn,aps,prc,superscriptaddress,showpacs,floatfix]{revtex4}
\usepackage{amssymb}
\usepackage{amsmath}
\usepackage{graphicx}
\usepackage[normalem]{ulem}
\usepackage{color}
\renewcommand\sout{\bgroup \color{red} \ULdepth=-.5ex \ULset}

\begin{document}

\title{Antimatter $^4_{\Lambda}$H Hypernucleus Production and the $^3_{\Lambda}$H/$^3$He Puzzle in Relativistic Heavy-Ion Collisions}
\author{Kai-Jia Sun}$  $
\affiliation{Department of Physics and Astronomy and Shanghai Key Laboratory for
Particle Physics and Cosmology, Shanghai Jiao Tong University, Shanghai 200240, China}
\author{Lie-Wen Chen \footnote{Corresponding author (email: lwchen$@$sjtu.edu.cn)}}
\affiliation{Department of Physics and Astronomy and Shanghai Key Laboratory for
Particle Physics and Cosmology, Shanghai Jiao Tong University, Shanghai 200240, China}
\affiliation{Center of Theoretical Nuclear Physics, National Laboratory of Heavy Ion
Accelerator, Lanzhou 730000, China}
\date{\today}

\begin{abstract}
We show that the measured yield ratio $^3_{\Lambda}$H/$^3$He
($^3_{\overline{\Lambda}}\overline{\text{H}}$/$^3\overline{\text{He}}$)
in Au+Au collisions at $\sqrt{s_{NN}}=200$ GeV and in Pb+Pb collisions
at $\sqrt{s_{NN}}=2.76$ TeV can be understood within a covariant coalescence
model if (anti-)$\Lambda$ particles freeze out earlier than (anti-)nucleons
but their relative freezeout time is closer at $\sqrt{s_{NN}}=2.76$ TeV
than at $\sqrt{s_{NN}}=200$ GeV.
The earlier (anti-)$\Lambda$ freezeout
can significantly enhance the yield of (anti)hypernucleus $^4_{\Lambda}$H
($^4_{\overline{\Lambda}}\overline{\text{H}}$), leading to that
$^4_{\overline{\Lambda}}\overline{\text{H}}$ has a comparable abundance with
$^4\overline{\text{He}}$ and thus provides an easily measured antimatter candidate
heavier than $^4\overline{\text{He}}$.
The future measurement
on $^4_{\Lambda}$H ($^4_{\overline{\Lambda}}\overline{\text{H}}$) would be very useful
to understand the (anti-)$\Lambda$ freezeout dynamics and the production mechanism of
(anti)hypernuclei in relativistic heavy-ion collisions.
\end{abstract}

\maketitle

\emph{1. Introduction.}---%
The recent observations of light antinuclei in relativistic heavy-ion
collisions at the Relativistic Heavy-Ion Collider (RHIC)~\cite{Abe09,Aga11} and
Large Hadron Collider (LHC)~\cite{Sha11,ALICE15} attract strong interest on the
study of antimatter~\cite{Ma12}, and verify the general principles of quantum field
theory which requires that each particle has its corresponding antiparticle and any physical
system has an antimatter analog with an identical mass (but the opposite charge).
These studies also provide the possibility in
laboratories to test the fundamental CPT theorem~\cite{Ada15}, to explore the
interactions between antimatter and antimatter~\cite{STAR15}, and to help hunting for antimatter
and dark matter in the Universe through cosmic radiation observations~\cite{Car14}.
The antihelium-4 ($^4\overline{\text{He}}$ or $\overline{\alpha}$) is
the heaviest antimatter nucleus observed so far~\cite{Aga11}, and it is of great interest to search
for antimatter nuclei heavier than $^4\overline{\text{He}}$ in heavy-ion collisions,
which is extremely useful to understand the production mechanism of heavier
antimatter~\cite{Gre96,SunKJ15}.

Collisions of heavy nuclei at high energies
also provide an abundant source of (anti)strangeness~\cite{Koc86} and a unique tool to
produce light (anti-)hypernulcei~\cite{Ko85}.
The STAR collaboration
at RHIC reported the observation of hypertriton ($^3_{\Lambda}$H)
and antihypertriton ($^3_{\overline{\Lambda}}\overline{\text{H}}$) in Au+Au
collisions at $\protect\sqrt{s_{NN}}=200$ GeV~\cite{Abe10}, and recently the ALICE
collaboration at LHC also reported the observation in Pb+Pb
collisions at $\protect\sqrt{s_{NN}}=2.76$ TeV~\cite{ALICE15H}.
The value of measured yield ratio $^3_{\Lambda}$H/$^3$He is $0.82\pm0.16(\text{stat.})\pm0.12(\text{syst.})$
for $0-80$\% centrality Au+Au
collisions at $\protect\sqrt{s_{NN}}=200$ GeV (RHIC)~\cite{Abe10} and
$0.47\pm0.10(\text{stat.})\pm0.13(\text{syst.})$ in central ($0-10$\% centrality) Pb+Pb
collisions at $\protect\sqrt{s_{NN}}=2.76$ TeV
(LHC)~\cite{ALICE15H}.
It is thus favored that the measured $^3_{\Lambda}$H/$^3$He ratio at RHIC is
higher than that at LHC, although they are compatible with a very small overlap
within the uncertainties by combing the statistical and systematic
uncertainties, i.e., $0.82\pm0.20$ at RHIC and $0.47\pm0.16$ at LHC.
The value of the $^3_{\Lambda}$H/$^3$He ratio for $0-80$\% centrality at RHIC (i.e., $0.82\pm0.20$)
is expected to be further enhanced for central collisions since the ALICE measurements indicate
that the $^3_{\Lambda}$H/$^3$He ratio in central Pb+Pb collisions is higher than that in
peripheral Pb+Pb collisions~\cite{ALICE15H}.
Similar conclusion is obtained for the
$^3_{\overline{\Lambda}}\overline{\text{H}}$/$^3\overline{\text{He}}$
ratio for which the measured value is $0.89\pm0.28(\text{stat.})\pm0.13(\text{syst.})$
for $0-80$\% centrality Au+Au collisions at $\protect\sqrt{s_{NN}}=200$ GeV~\cite{Abe10} and
$0.42\pm0.10(\text{stat.})\pm0.13(\text{syst.})$ in central ($0-10$\% centrality) Pb+Pb
collisions at $\protect\sqrt{s_{NN}}=2.76$ TeV~\cite{ALICE15H}.
As shown in Ref.~\cite{ALICE15H},
the conventional (statistical) thermal models~\cite{And11,Cle11,Pal13,Pet13}
failed to describe the RHIC ratio, although some of them~\cite{And11,Cle11,Pal13}
successfully described the LHC ratio. The thermal model with a multi-freezeout
configuration~\cite{Cha14} reasonably described the ratio at RHIC but failed at LHC,
and so did the parton and hadron cascade plus dynamically constrained
phase-space coalescence %(PACIAE+DCPC)
model~\cite{ChenG13,She15}.
The dynamical~\cite{Zha10} and simple~\cite{Xue12,Sha15} coalescence models
described marginally the ratio at RHIC but no results are available at LHC.
Therefore, the measured $^3_{\Lambda}$H/$^3$He and
$^3_{\overline{\Lambda}}\overline{\text{H}}$/$^3\overline{\text{He}}$ ratios
challenge all theoretical calculations performed so far and call for novel
mechanisms for (anti-)hypernuclei production in these collisions.

Since hyperons have quite different interactions compared with
nucleons~\cite{Bot12}, they are expected to have different freezeout
dynamics in heavy-ion collisions, which will lead to distinct features for
the production of light (anti)hypernuclei compared with that of light
normal (anti)nuclei.
In this work,
we show that the covariant coalescence model can naturally reproduce
the measured $^3_\Lambda$H/$^3$He ($^3_{\overline{\Lambda}}\overline{\text{H}}$/$^3\overline{\text{He}}$)
at both RHIC and LHC if (anti-)$\Lambda$ particles freeze out
earlier than (anti)nucleons but their relative freezeout time is closer
at LHC than at RHIC.
The earlier anti-$\Lambda$ ($\overline{\Lambda}$) freezeout
leads to that the heavier antihypernucleus
$^4_{\overline{\Lambda}}\overline{\text{H}}$ has a comparable yield with
$^4\overline{\text{He}}$ and thus provides an easily measured candidate
for antimatter heavier than $^4\overline{\text{He}}$.

\emph{2. Covariant coalescence model.}---%
We use the covariant coalescence model~\cite{Dov91} for the production of
light clusters in heavy-ion collisions.
The main feature of
the coalescence model~\cite{But61,Sat81,Cse86} is that the coalescence probability
depends on the details of the phase space structure of the constituent particles at
freezeout as well as the statistical weight and internal structure (wave function)
of the coalesced cluster, and these details are of no relevance in the thermal
model~\cite{Cle91,Bra95,And11,Cle11,Ste12} of cluster creation.

The phase space configuration of the constituent particles at freezeout is a
basic ingredient in the coalescence model, and in principe it can be obtained
dynamically from transport model simulations for heavy-ion collisions
(see, e.g., Refs.~\cite{Mat97,ChenLW03,ChenLW06,Oh09,Zhu15}).
For the particle production at mid-rapidity
in central heavy-ion collisions at
RHIC and LHC considered here, for simplicity, we assume a boost-invariant
longitudinal expansion for the constituent particles which are emitted from a
freezeout hypersurface $\Sigma^\mu$, and the Lorentz invariant one-particle
momentum distribution is then given by
\begin{eqnarray}
E\frac{d^3N}{d^3p}=\frac{d^3N}{p_Tdp_T d\phi_p dy } = \int\limits_{\Sigma^\mu}\text{d}\sigma _{\mu} p^\mu f(x,p) = \int d^4x S(x,p),
\end{eqnarray}
where $\sigma_\mu$ denotes the normal vector of hypersurface $\Sigma^\mu$ and $p^\mu$ is the
four-momentum of the emitted particle. The emission function $S(x,p)$ can be expressed by
\begin{eqnarray}
S(x,p)d^4x =  m_T\cosh(\eta-y)f(x,p)J(\tau) \tau \text{d}\tau\text{d}\eta r \text{d}r \text{d}\phi_s,
\end{eqnarray}
where we use longitudinal proper time $\tau = \sqrt{t^2-z^2}$, spacetime rapidity
$\eta = \frac{1}{2} \text{ln}\frac{t-z}{t+z}$, cylindrical coordinates ($r$, $\phi_s$),
rapidity $y=\frac{1}{2}\ln (\frac{E+p_z}{E-p_z})$, transverse momentum ($p_T,\phi_p$),
and transverse mass $m_T=\sqrt{m^2+p_T^2}$.
The statistical distribution function $f(x,p)$ is given
by $f(x,p)=g(2\pi)^{-3}[\exp(p^{\mu}u_{\mu}/kT)/\xi \pm 1]^{-1}$
with $g$ being spin degeneracy factor, $\xi$ the fugacity, $u_{\mu}$ the
four-velocity of a fluid element in the fireball, $T$ the local temperature, and
$p^{\mu} u_{\mu}=m_T \cosh\rho \cosh(\eta -y)-p_T \sinh\rho \cos(\phi_p -\phi_s)$
the energy in local rest frame of the fluid.
Following Ref.~\cite{Ret04},
we assume the freezeout time follows a Gaussian distribution
$J(\tau)=\frac{1}{\Delta \tau \sqrt{2\pi}}\exp(-\frac{(\tau-\tau_0)^2}{2(\Delta \tau)^2})$
with a mean value $\tau_0$ and a dispersion $\Delta \tau$,
and the transverse rapidity distribution of the fluid element in the fireball is
parameterized as $\rho=\rho_0 r/R_0$ with $\rho_0$ being the maximum transverse
rapidity and $R_0$ the transverse radius of the fireball.
The phase space freezeout configuration of the constituent particles
is thus determined by six parameters, i.e., $T$, $\rho_0$, $R_0$, $\tau_0$,
$\Delta \tau$ and $\xi$.

The cluster production probability is determined by
the overlap of the cluster Wigner function with the constituent
particle phase-space distribution at freezeout. If $M$ particles
are coalesced into a cluster, the invariant
momentum distribution of the cluster can be obtained as
\begin{eqnarray}
E\frac{d^3N_c}{d^3P}&=&Eg_c\int  \bigg(\prod_{i=1}^{M} \frac{d^3p_i }{E_i}d^4x_iS(x_i,p_i)\bigg)\times \notag \\
&&\rho_c^W(x_1,...,x_M;p_1,...,p_M)\delta^3(\mathbf{P}-\sum_{i=1}^M\mathbf{p_i}),
\label{Eq:Coal}
\end{eqnarray}
where $N_c$ is the cluster multiplicity, $E$ ($\mathbf{P}$) is its
energy (momentum), $g_c$ is the coalescence factor, and
$\rho_c^W$ is the Wigner function.
In this work, the harmonic oscillator wave functions are assumed for all the
clusters in the rest frame except the (anti)deutrons for which
the Hulth\'{e}n wave function is used (see, e.g.,
Refs.~\cite{Mat97,ChenLW03}), and so the cluster Wigner functions
and root-mean-square radii $r_{\text{rms}}$ can
be obtained analytically.
The details about how to calculate the integral (\ref{Eq:Coal}) can
be found in Ref.~\cite{SunKJ15}. It should be stressed that since
the constituent particles may have different freezeout time, the
particles that freeze out earlier are allowed to propagate freely
until the time when the last particles in the cluster freezes out
to make the coalescence at equal time~\cite{Mat97,ChenLW06,SunKJ15}.

\emph{3. Production of (anti)hypertriton.}---%
The coalescence factor is given by $g_c = \frac{2j+1}{2^N}$~\cite{Sat81}
with $j$ the spin and $N$ the nucleon number of the nucleus.
For d, $^3$He, $^3_{\Lambda}$H, $^4$He and $^4_{\Lambda}$H that we focus on here, their spins
are $1$, $1/2$, $1/2$, $0$ and $0$, respectively, and their $r_{\text{rms}}$ which are
directly related to their Wigner functions~\cite{SunKJ15}, are $1.96$ fm,
$1.76$ fm, $4.9$ fm, $1.45$ fm and $2.0$ fm, respectively~\cite{Rop09,Nem00}. The anti(hyper)nuclei
are assumed to have the same $j$ and $r_{\text{rms}}$ as their corresponding (hyper)nuclei.

\begin{figure}
\includegraphics[scale=0.36]{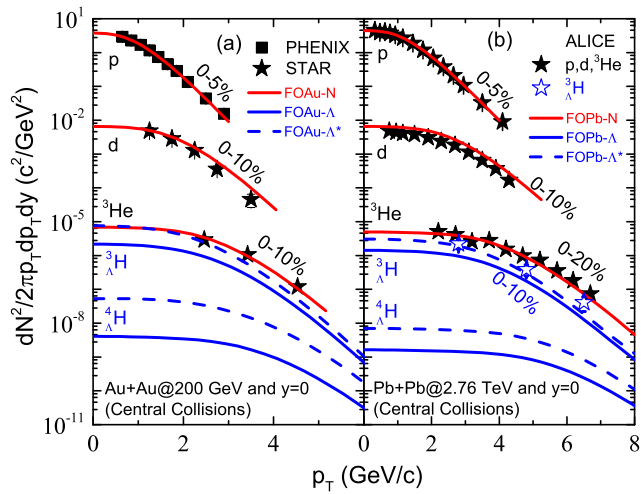}
\caption{Transverse momentum distributions of p, d, $^3$He, $^3_{\Lambda}$H and
$^4_{\Lambda}$H at midrapidity in central collisions of Au+Au at $\sqrt{s_{NN}}=200$ GeV
(a) and Pb+Pb at $\sqrt{s_{NN}}=2.76$ TeV (b)
predicted by coalescence model with various freezeout configurations.
For Au+Au collisions, the data of proton is taken from the PHENIX~\cite{Adl04} and
those of d and $^3$He are taken from STAR~\cite{Abe09}. The data of p, d, $^3$He
and $^3_\Lambda\text{H}$ for Pb+Pb collisions are taken from
ALICE~\cite{ALICE12,ALICE15H,ALICE15}.}
\label{FigSpectNucl}
\end{figure}

Following Ref.~\cite{SunKJ15},
the proton (p) freezeout parameters $T$ and $\rho_0$ can
be extracted from fitting the p spectrum, and the $R_0$, $\tau_0$,
$\Delta \tau$ and $\xi_p$ can be obtained by further fitting the spectra of d and
$^3$He in the coalescence model. For central Au+Au collisions at $\sqrt{s_{NN}}=200$ GeV,
we obtain $T=111.6$ MeV, $\rho_0 = 0.98 $, $R_0 = 15.6$ fm, $\tau_0 = 10.6$ fm/c,
$\Delta \tau = 3.5$ fm/c and $\xi_p = 10.5$
by fitting the p spectrum from PHENIX~\cite{Adl04} and
the spectra of d and $^3$He from STAR~\cite{Abe09}, and
for antiprotons ($\overline p$), we assume they have
the same freezeout as protons except the fugacity is
reduced to $\xi_{\overline p}=7.84$ to describe the measured
$\overline {p}/p=0.75$~\cite{Adl04}.
For central Pb+Pb collisions at $\protect\sqrt{s_{NN}}=2.76$ TeV, we obtain
$T=121.1$ MeV, $\rho_0 = 1.215$, $R_0=19.7$ fm, $\tau_0=15.5$ fm/c,
$\Delta \tau=1.0$ fm/c and $\xi_p = 3.72$ by fitting the measured spectra
of p, d, $^3\text{He}$ from ALICE~\cite{Abe09,ALICE12,ALICE15}, and
the antiprotons are assumed to have the same freezeout parameters as
protons since the ${\overline p}/p$ is close to unity at LHC.
For neutrons (n) (antineutrons ($\overline {\text{n}}$)), we take
their freezeout parameters as those of p's ($\overline {\rm p}$'s) since the
isospin chemical potential at freezeout is small at RHIC and LHC~\cite{And11}.
The p freezeout parameters at RHIC (denoted by FOAu-N) and LHC (denoted by FOPb-N)
are summarized in Table~\ref{TabParam}, and the freezeout hypersurface at LHC is seen
to have larger $T$, $\rho_0$, $R_0$ and $\tau_0$ but smaller
$\Delta \tau$ and $\xi_p$.
In Fig.~\ref{FigSpectNucl},
the experimental data are compared with the calculated results
for the spectra of p, d and $^3$He with FOAu-N and FOPb-N, and one can see that
the coalescence model describes well the measured spectra.
Table~\ref{LightNuclYield} lists
the $p_T$-integrated yield in the midrapidity region ($-0.5\leq$y$\leq 0.5$) (i.e., $dN/dy$)
for p ($\overline{\text{p}}$), d ($\overline{\text{d}}$), $^3$He ($^3\overline{\text{He}}$)
and $^4$He ($^4\overline{\text{He}}$) with FOAu-N and FOPb-N, and it is seen that the $dN/dy$
values of d, $^3$He and $^4$He ($\overline{\text{d}}$,
$^3\overline{\text{He}}$ and $^4\overline{\text{He}}$) at LHC are
roughly two (four) times as large as those at RHIC.

\begin{table}
\caption{Parameters of various freezeout configurations for (anti)nucleons and
(anti-)$\Lambda$ particles at midrapidity in central collisions of Au+Au at
$\sqrt{s_{NN}}=200$ GeV (FOAu) and Pb+Pb at $\sqrt{s_{NN}}=2.76$ TeV (FOPb).
The unit of $\tau_0$ and $\Delta \tau$ is fm/c. $\xi$ and $\bar{\xi}$ denote the
fugacity of particles and antiparticles, respectively.}
\begin{tabular}{c|c|c|c|c|c|c|c}
        \hline \hline
          & $T$ (MeV) & $\rho_0$ & $R_0$ (fm) & $ \tau_0$ & $ \Delta \tau$ & $\xi$ & $\bar{\xi}$ \\
         \hline
        FOAu-N~ & 111.6  & 0.980 & 15.6 & 10.6 & 3.5 & 10.5  &7.84 \\
        FOPb-N~ & 121.1  & 1.215 & 19.7 & 15.5& 1.0 & 3.72 &  3.72 \\

         \hline
        FOAu-$\Lambda$~& 111.6  & 0.980 & 15.6 & 10.6 & 3.5 & 42.8 & 35.1  \\
        FOPb-$\Lambda$~& 121.1  & 1.215 & 19.7 & 15.5& 1.0 & 9.54 & 9.54 \\

         \hline
        FOAu-$\Lambda^*$ & 126.0  & 0.890 & 11.1 & 7.54& 3.5 & 35.1 & 28.8 \\
        FOPb-$\Lambda^*$ & 123.4  & 1.171 & 16.7 & 13.1& 1.0 & 13.6 & 13.6 \\
        \hline  \hline
\end{tabular}
\label{TabParam}
\end{table}

\begin{table}
\scriptsize
\caption{$dN/dy$ at midrapidity of light (anti)(hyper)hypernuclei for
various freezeout configurations in central collisions of Au+Au at
$\sqrt{s_{NN}}=200$ GeV (FOAu) and Pb+Pb at $\sqrt{s_{NN}}=2.76$ TeV (FOPb).}
\begin{tabular}{c|c|c|c|c}
        \hline \hline
           & p($\overline{\text{p}}$ ) & d($\overline{\text{d}}$) & $^3$He( $^3\overline{\text{He}}$) & $^4$He($\overline{^4\text{He}}$) \\
         \hline
        FOAu-N &16.1(12.1) &7.49(4.21)$\times 10^{-2}$& 14.9(6.29)$\times 10^{-5}$& 15.4(4.88)$\times 10^{-8}$ \\

        FOPb-N &33.5(33.5) &15.0(15.0)$\times 10^{-2}$& 2.36(2.36)$\times 10^{-4}$& 2.20(2.20)$\times 10^{-7}$ \\

        \hline \hline
           & $\Lambda$($\overline{\Lambda}$) & &$^3_{\Lambda}$H($^3_{\overline{\Lambda}}\overline{\text{H}}$) & $^4_{\Lambda}$H($^4_{\overline{\Lambda}}\overline{\text{H}}$)     \\
         \hline
        FOAu-$\Lambda$ ~& 17.0(14.0) &  &4.26(1.96)$\times 10^{-5}$ & 14.8(5.12)$\times 10^{-8}$   \\
        FOPb-$\Lambda$ ~& 24.9(24.9) & & 5.72(5.72)$\times 10^{-5}$& 1.36(1.36)$\times 10^{-7}$      \\

         \hline
        FOAu-$\Lambda^*$ & 18.8(15.4) &  & 12.3(5.65)$\times 10^{-5}$& 15.7(5.43)$\times 10^{-7}$     \\
        FOPb-$\Lambda^*$ & 25.9(25.9) & & 1.12(1.12)$\times 10^{-4}$& 5.43(5.43)$\times 10^{-7}$      \\

          \hline \hline
\end{tabular}
\label{LightNuclYield}
\end{table}

\begin{figure}
\includegraphics[scale=0.36]{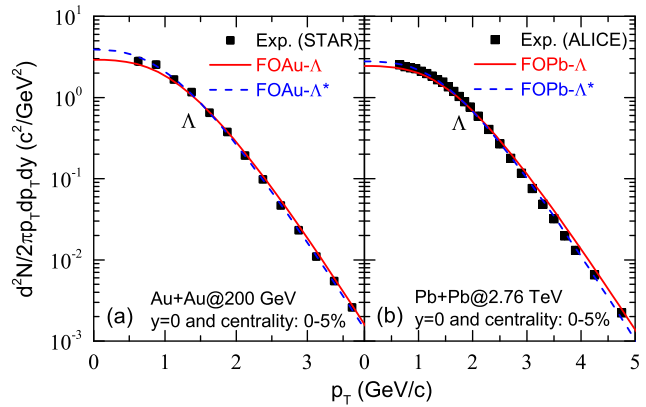}
\caption{Transverse momentum distribution of $\Lambda$'s in central collisions of Au+Au at
$\sqrt{s_{NN}}=200$ GeV (a) and Pb+Pb at $\sqrt{s_{NN}}=2.76$ TeV (b)
from coalescence model calculations with various freezeout configurations.
The experimental data are taken from STAR~\cite{Aga04} for the Au+Au collisions and
from ALICE~\cite{ALICE13} for the Pb+Pb collisions.}
\label{FigSpectLam}
\end{figure}

For $\Lambda$ particles, we first assume they have the same freezeout configuration
as nucleons except that the $\Lambda$ fugacity becomes $\xi_\Lambda=42.8~(9.54)$
at RHIC (LHC) by fitting the experimental $\Lambda$
spectra~\cite{Aga04,ALICE13} as shown in Fig.~\ref{FigSpectLam} by solid lines.
The $\overline{\Lambda}$ particles are assumed to have the same freezeout
parameters as $\Lambda$ particles except the fugacity at RHIC is reduced to
$\xi_{\overline \Lambda}=35.1$ to describe the measured
$\overline \Lambda/\Lambda=0.82$~\cite{Aga04}.
The (anti-)$\Lambda$ freezeout parameters are listed as FOAu-$\Lambda$ (FOPb-$\Lambda$)
in Table~\ref{TabParam} for the central Au+Au (Pb+Pb) collisions.
With FOAu-$\Lambda$ and FOPb-$\Lambda$ (together with FOAu-N and FOPb-N),
the spectra of $^3_{\Lambda}$H and $^4_{\Lambda}$H can then be calculated
using the coalescence model, and the results are shown in Fig.~\ref{FigSpectNucl}.
The $dN/dy$ values for $\Lambda$ ($\overline{\Lambda}$),
$^3_{\Lambda}$H ($^3_{\overline{\Lambda}}\overline{\text{H}}$) and
$^4_{\Lambda}$H ($^4_{\overline{\Lambda}}\overline{\text{H}}$)
are summarized in Table~\ref{LightNuclYield}, and
the resulting $^3_{\Lambda}$H/$^3$He is about $0.29$ ($0.24$)
at RHIC (LHC) with FOAu-$\Lambda$ (FOPb-$\Lambda$), which significantly
underestimates the measured values from STAR~\cite{Abe10} and
ALICE~\cite{ALICE15H},
i.e., $0.82\pm0.16(\text{stat.})\pm0.12(\text{syst.})$
for $0-80$\% centrality Au+Au
collisions at $\protect\sqrt{s_{NN}}=200$ GeV~\cite{Abe10} and
$0.47\pm0.10(\text{stat.})\pm0.13(\text{syst.})$ in central ($0-10$\% centrality) Pb+Pb
collisions at $\protect\sqrt{s_{NN}}=2.76$
TeV~\cite{ALICE15H}.
The predicted ratio $^3_{\overline{\Lambda}}\overline{\text{H}}$/$^3\overline{\text{He}}$
is about $0.31$ ($0.24$) at RHIC (LHC) with FOAu-$\Lambda$ (FOPb-$\Lambda$),
again significantly underestimating the measured values, i.e.,
$0.89\pm 0.28(\text{stat.})\pm 0.13(\text{syst.})$ from STAR~\cite{Abe10} and
$0.42\pm 0.10(\text{stat.})\pm 0.13(\text{syst.})$ from ALICE~\cite{ALICE15H}.
In addition, the predicted $^3_{\Lambda}$H spectrum with FOPb-$\Lambda$ is seen
to underestimate the recently measured spectrum by ALICE~\cite{ALICE15H}.

To understand the disagreement of $^3_{\Lambda}$H/$^3$He
($^3_{\overline{\Lambda}}\overline{\text{H}}$/$^3\overline{\text{He}}$)
and the $^3_{\Lambda}$H spectrum between the predictions and the measurements,
we extract the $\Lambda$ freezeout parameters $T$ and $\rho_0$
by directly fitting the measured $\Lambda$ spectra~\cite{Aga04,ALICE13} as shown in
Fig~\ref{FigSpectLam} by dashed lines, and we obtain $T =126~(123.4)$ MeV and
$\rho_0 = 0.89~(1.171)$ for Au+Au (Pb+Pb) collisions, which better
describe the data than FOAu-$\Lambda$ (FOPb-$\Lambda$).
The $\Lambda$ particles thus have a higher freezeout temperature than nucleons,
especially at RHIC, implying an earlier freezeout for $\Lambda $
particles than for nucleons, which is consistent with the empirical picture
that the strange baryons usually freeze out earlier than the nonstrange baryons
due to their relatively smaller interaction cross sections.
The earlier $\Lambda $ freezeout is also supported
by the investigation on strangeness production~\cite{Cha14,He10,Bug13,Cha13,Cha15}
as well as the microscopic transport model simulations~\cite{Bra05}.

An earlier $\Lambda $ freezeout means the $\Lambda $ particles can pick up
unfrozen-out nucleons to form light hypernuclei, and this implies the nucleons
coalesced into light hypernuclei also have an earlier freezeout time
than those coalesced into normal light nuclei.
To consider this effect, for the coalescence production of
light hypernuclei, we reduce the $\tau_0$ and $R_0$ simultaneously but increase
the $\xi$ to fit the $\Lambda $ and p spectra. In this way, the earlier
freezeout increases the phase space density of $\Lambda $, p
and n, and thus the $^3_\Lambda$H production rate.
To fit the measured central value $0.82$ ($0.47$) of the
$^3_\Lambda$H/$^3$He ratio at RHIC (LHC), we find the $R_0$ and $\tau_0$ need to be
reduced to $71\%$ ($85\%$) of their values in FOAu-$\Lambda$ (FOPb-$\Lambda$),
i.e., $R_0 = 11.1~(16.7)$ fm and $\tau_0 = 7.54~(13.1)$ fm/c, if we
fix $T =126~(123.4)$ MeV, $\rho_0 = 0.89~(1.171)$ and $\Delta \tau = 3.5~(1.0)$ fm/c.
These new freezeout configurations are denoted as FOAu-$\Lambda^*$ and FOPb-$\Lambda^*$
in Table~\ref{TabParam}.
It is interesting to note that the $\Lambda $ freezeout temperature is slightly
higher at RHIC than at LHC.
For FOAu-$\Lambda^*$ and FOPb-$\Lambda^*$,
we have neglected final state interactions of the produced $^3_{\Lambda}$H
during the last $2\sim 3$ fm/c time interval when some nucleons have not yet
frozen out, and probably this can be justified from the transport model
study which indicates including the final state interactions changes the
deuteron yield by only about $20\%$ at RHIC~\cite{Oh09}.
On the other hand,
since $^3_\Lambda$H is an even more loosely bound system than deuteron
(Note the total binding energy of $^3_\Lambda$H is $2.354$ MeV with the $\Lambda $
separation energy of only about $0.13$ MeV~\cite{Jur73}, and the total binding energy of
deuteron is $2.224$ MeV~\cite{Wan12}), the effects of the final state interactions on
$^3_\Lambda$H yield are thus expected to be stronger than
that on deuteron.
For $^4_\Lambda$H, the total binding energy is $10.601$ MeV
with the $\Lambda $ separation energy of $2.12$ MeV~\cite{Ess15}, and the effects of
the final state interactions are thus expected to be similar with the
case of deuteron. The quantitative information on the final interaction
effects needs a complicated transport model simulations.
The stronger final state interaction (destruction) of $^3_\Lambda$H
implies the $\Lambda $ particles need an even earlier freezeout than
that obtained above, and the effects of an earlier $\Lambda $ freezeout
in the present work are thus considered to be conservative estimate.

The predicted spectra of $^3_{\Lambda}$H and $^4_{\Lambda}$H with FOAu-$\Lambda^*$
and FOPb-$\Lambda^*$ are shown in Fig.~\ref{FigSpectNucl} by dashed lines, and one
can see that compared with FOAu-$\Lambda$ and FOPb-$\Lambda$,
FOAu-$\Lambda^*$ and FOPb-$\Lambda^*$ significantly enhance the production of $^3_{\Lambda}$H
and $^4_{\Lambda}$H and now the $^3_{\Lambda}$H spectra measured by ALICE~\cite{ALICE15H}
can be reasonably described by FOPb-$\Lambda^*$.
The $dN/dy$ values for $\Lambda$ ($\overline{\Lambda}$),
$^3_{\Lambda}$H ($^3_{\overline{\Lambda}}\overline{\text{H}}$) and
$^4_{\Lambda}$H ($^4_{\overline{\Lambda}}\overline{\text{H}}$)
with FOAu-$\Lambda^*$ and FOPb-$\Lambda^*$ are listed in
Table~\ref{LightNuclYield}.
The small difference for the $dN/dy$ of (anti-)$\Lambda$ between FOAu-$\Lambda^*$
(FOPb-$\Lambda^*$) and FOAu-$\Lambda$ (FOPb-$\Lambda$) is due to the slight variation
of the $\Lambda$ spectra from different fits as shown in Fig~\ref{FigSpectLam}.
The calculated $^3_{\Lambda}$H/$^3$He and
$^3_{\overline{\Lambda}}\overline{\text{H}}$/$^3\overline{\text{He}}$ ratios
are, respectively, about $0.83$ ($0.47$) and $0.91$ ($0.47$) at RHIC (LHC) with
FOAu-$\Lambda^*$ (FOPb-$\Lambda^*$), nicely reproducing the measured central
values.
Therefore, our results suggest
that the (anti-)$\Lambda$ particles may freeze out earlier than (anti)nucleons but
their relative freezeout time is closer at LHC than at RHIC.
It is interesting to see
that the $\Lambda$ and nucleon freezeout parameters seem to come close to
each other as the energy increases and this is understandable since a higher
colliding energy generally leads to a longer-lived hadronic fireball where
the $\Lambda$'s and nucleons will experience more collisions.

The ratio $S_3 = ^3_{\Lambda}$H/($^3$He$\times \Lambda/$p) was first suggested
in Ref.~\cite{Arm04} in the expectation that dividing the strange to
nonstrange baryon yield should result in a value near unity in a naive
coalescence model. It was also argued~\cite{Zha10} to be a good
representation of the local correlation between baryon number and
strangeness~\cite{Koc05}, and thus should be a valuable probe for
the onset of deconfinement in relativistic heavy-ion collisions.
The $S_3$ was measured
to be $1.08 \pm 0.22$ for $0$-$80\%$ centrality Au+Au collisions~\cite{Abe10}
and $0.60\pm 0.13(\text{stat.})\pm0.21(\text{syst.})$
for central ($0$-$10\%$ centrality) Pb+Pb collisions~\cite{ALICE15H}.
For central collisions considered here, the $S_3$ for Au+Au (Pb+Pb)
collisions is $0.27$ ($0.33$) with FOAu-$\Lambda$ (FOPb-$\Lambda$), while it
increases to $0.71$ ($0.61$) with FOAu-$\Lambda^*$ (FOPb-$\Lambda^*$).
The FOAu-$\Lambda$ (FOPb-$\Lambda$) thus
significantly underestimates the measured $S_3$ for Au+Au (Pb+Pb)
collisions. While FOPb-$\Lambda^*$ nicely reproduces the measured
$S_3$ for Pb+Pb collisions, the FOAu-$\Lambda^*$ still
underestimates the measured $S_3$ for Au+Au collisions.
It should be noted
that while there is negligible feed-down from heavier states into
$^3_{\Lambda}$H and $^3$He, the $\Lambda$ and p are significantly influenced
by feed-down from decays of excited baryonic states.
In the coalescence model calculations, the $\Lambda$ and p from the
short-lived strong decays are included
since they appear in the fireball, while those from the other
long-lived decays are excluded since they are out of the fireball.
In the calculation of the $S_3$ for Au+Au collisions,
we use the p spectrum from PHENIX~\cite{Adl04} which is corrected by excluding
the contribution from the long-lived weak decays. We note that including $40\%$
contribution from weak decays to the p yield leads to $S_3 = 0.994$ for Au+Au
collisions with FOPb-$\Lambda^*$, consistent with the measured value from STAR.

\begin{figure}
\includegraphics[scale=0.32]{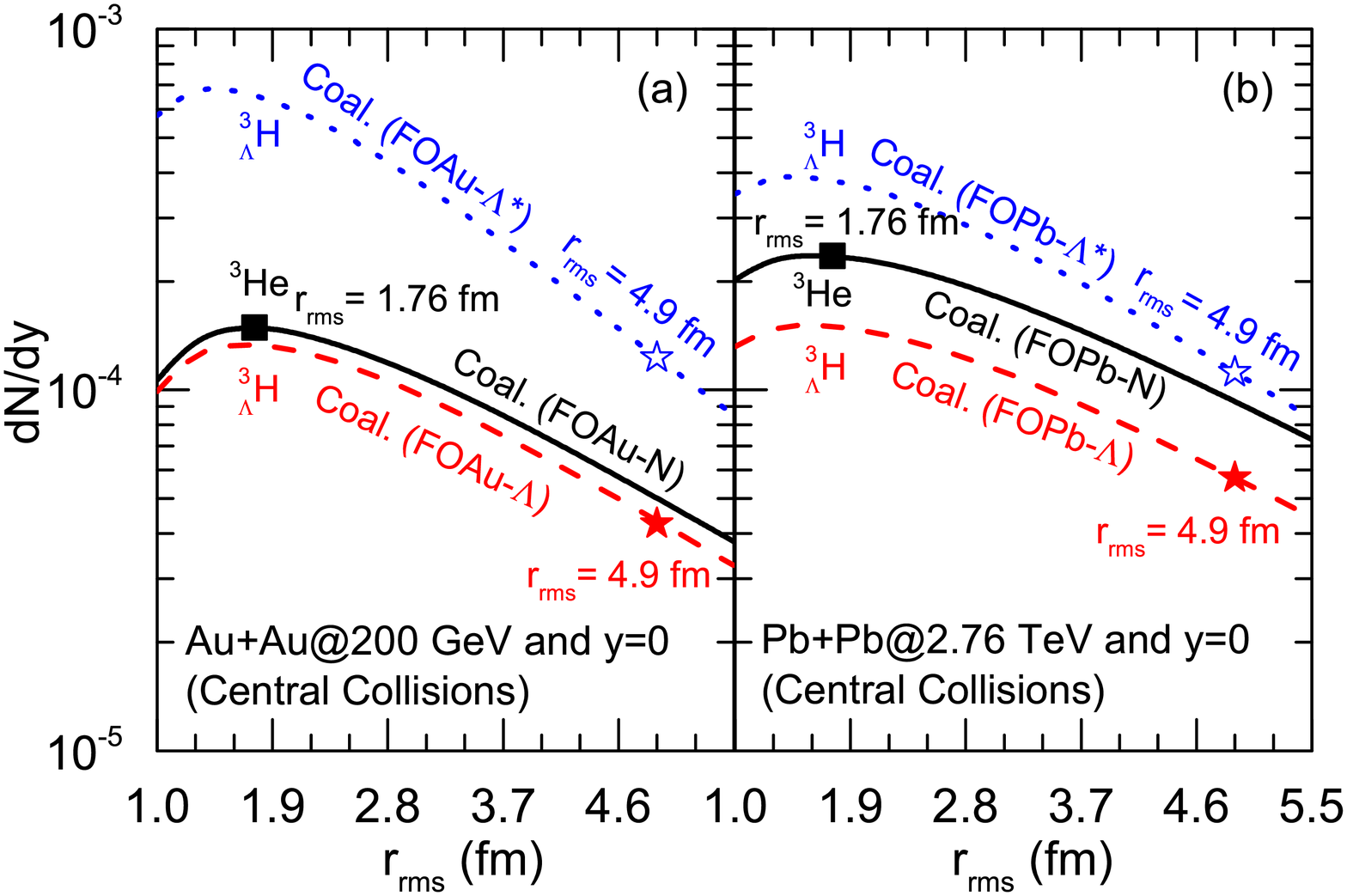}
\caption{The predicted $dN/dy$ of $^3_\Lambda\text{H}$ and $^3\text{He}$ at midrapidity
as a function of their root-mean-square radii $r_{\rm{rms}}$ in central collisions of Au+Au at
$\sqrt{s_{NN}}=200$ GeV (a) and Pb+Pb at $\sqrt{s_{NN}}=2.76$ TeV (b)
from the coalescence model with various freezeout configurations.
The stars (squares) indicate the empirical size values
$r_{\rm {rms}}=4.9~(1.76)$ fm for $^3_\Lambda\text{H}$ ($^3\text{He}$).}
\label{Fig:rms}
\end{figure}

It should be pointed out that
although the $\Lambda$'s and nucleons are assumed to have the same freezeout configuration, the $S_3$ is
still significantly less than unity (e.g., $S_3 = 0.27$ ($0.32$) for FOAu-$\Lambda$
(FOPb-$\Lambda$)). This is mainly due to the much larger size of $^3_{\Lambda}$H than
that of $^3$He, as suggested first in Ref.~\cite{Arm04}.
To see this more clearly,
we show in Fig.~\ref{Fig:rms} the predicted $dN/dy$ of $^3_\Lambda\text{H}$ and $^3\text{He}$
as a function of their root-mean-square radii in central collisions of Au+Au at
$\sqrt{s_{NN}}=200$ GeV and Pb+Pb at $\sqrt{s_{NN}}=2.76$ TeV
from the coalescence model with various freezeout configurations.
The empirical size values, i.e., $r_{\rm {rms}}=4.9$ fm for $^3_\Lambda\text{H}$
and $r_{\rm {rms}}=1.76$ fm for $^3\text{He}$ are also indicated in Fig.~\ref{Fig:rms}.
It is seen that, because of the finite size cut off effect of the fireball in the spatial
part integration of Eq.~(\ref{Eq:Coal}), the $dN/dy$ decrease with $r_{\rm {rms}}$
when $r_{\rm {rms}}$ is larger than about $1.6$ fm.
Furthermore, it is interesting to see that
the $dN/dy$ exhibits a stronger $r_{\rm {rms}}$ dependence at RHIC than that at LHC,
and this is mainly due to the fact that the freezeout volume ($\pi R_0^2\tau_0$) is
smaller at RHIC.
Compared with the thermal model, the coalescence model thus has a distinct feature
that the cluster yield depends on the cluster size, as mentioned earlier.
Assuming $^3_{\Lambda}$H has a
same $r_{\text{rms}}$ as $^3$He, i.e., $r_{\text{rms}} = 1.76$ fm, we find that the $S_3$
values for both Au+Au (with FOAu-$\Lambda$) and Pb+Pb (with FOPb-$\Lambda$) collisions
are drastically enhanced to about $0.85$, and further to unity if the $\Lambda$'s and
nucleons are assumed to have equal mass, as expected from the naive coalescence model.

\emph{4. Production of $^4_\Lambda $H ($^4_{\overline{\Lambda}}\overline{\text{H}}$).}---%
The $^4_\Lambda $H is a well-researched hypernucleus with lifetime of
$192_{-18}^{+20}$ ps~\cite{Rap14} and mass
$M(^4_\Lambda\text{H}) = 3922.484\pm 0.01(\text{stat.}) \pm 0.09(\text{syst.})$ MeV~\cite{Ess15}
(Note the mass of $^4\text{He}$ is $M(^4\text{He}) = 3727.379$ MeV).
The $^4_\Lambda $H can be
identified through the $^4\text{He}$-$\pi^-$ invariant mass spectrum from the
decay $^4_\Lambda \text{H}\rightarrow^4\text{He}+\pi^-$ with branching ratio of about
$50\%$~\cite{Kum95,Out98}.

\begin{figure}
\includegraphics[scale=0.28]{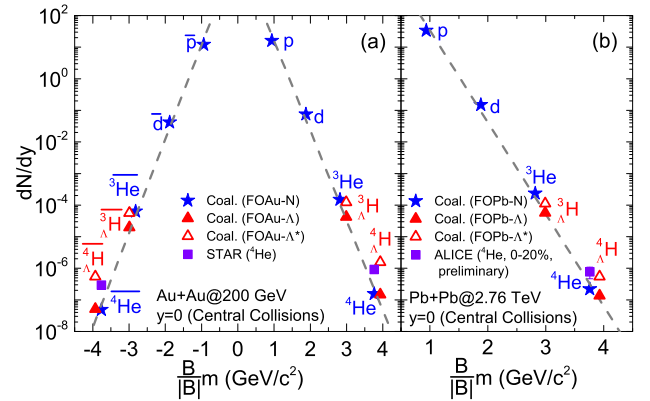}
\caption{The predicted $dN/dy$ of light (anti)(hyper)nuclei at midrapidity
as a function of $\frac{B}{|B|}m$ in central collisions of Au+Au at
$\sqrt{s_{NN}}=200$ GeV (a) and Pb+Pb at $\sqrt{s_{NN}}=2.76$ TeV (b)
from the coalescence model with various freezeout configurations.
The squares in (a) represent the results for $^4\text{He}$ and
$^4\overline{\text{He}}$ in the Au+Au collisions by considering the
binding energy effects to fit STAR data~\cite{Aga11,SunKJ15} while the square in (b) is the
preliminary result for $^4\text{He}$ in $0-20\%$ centrality Pb+Pb collisions
from ALICE measurement~\cite{Sha16}.}
\label{Fig:dndy}
\end{figure}

Shown in Fig.~\ref{Fig:dndy} are the predicted $dN/dy$ of light (anti)(hyper)nuclei
as a function of $\frac{B}{|B|}m$, where $B$ is the baryon number of light
clusters and $m$ is the corresponding mass, in central collisions of Au+Au at
$\sqrt{s_{NN}}=200$ GeV and Pb+Pb at $\sqrt{s_{NN}}=2.76$ TeV from the coalescence
model with various freezeout configurations.
For the Pb+Pb collisions,
we only show the results of (hyper)nuclei in Fig.~\ref{Fig:dndy} because the
results of anti-(hyper)nuclei are the same as those of their corresponding
(hyper)nuclei since the antiprotons (and anti-$\Lambda$'s) are assumed to have the same
freezeout configuration as their corresponding particles in central Pb+Pb collisions
at $\sqrt{s_{NN}}=2.76$ TeV.
Also included in Fig.~\ref{Fig:dndy} are
the preliminary result for $dN/dy$ (i.e., $7.8\pm$ 3.1$\times 10^{-7}$) of $^4\text{He}$
in central (0-20$\%$) Pb+Pb collisions
at $\sqrt{s_{NN}}=2.76$ TeV recently measured by ALICE~\cite{Sha16}
as well as the results for $dN/dy$ of $^4\text{He}$ and
$^4\overline{\text{He}}$ in central Au+Au collisions obtained from the
coalescence model by considering the binding energy effects to fit the
STAR data (for details see Ref~\cite{SunKJ15}).
Since FOAu-N and FOPb-N describe well the spectra of p, d and
$^3\text{He}$ as shown in Fig.~\ref{FigSpectNucl}, the predicted $dN/dy$ of
p ($\overline{\text{p}}$), d ($\overline{\text{d}}$) and $^3$He ($^3\overline{\text{He}}$)
in Fig.~\ref{Fig:dndy} are expected to give good estimates of the
experimental data on $dN/dy$.
On the other hand, FOAu-N and FOPb-N significantly underestimate the $dN/dy$ of
$^4\text{He}$ and $^4\overline{\text{He}}$, and these discrepancies can be
fixed by considering the effects of the large binding energy of $^4\text{He}$
and $^4\overline{\text{He}}$~\cite{SunKJ15}.

Furthermore, it is seen from
Fig.~\ref{Fig:dndy} that compared with FOAu-$\Lambda$ (FOPb-$\Lambda$),
FOAu-$\Lambda^*$  (FOPb-$\Lambda^*$) significantly enhances the $dN/dy$ of
(anti-)$^3_\Lambda\text{H}$ and (anti-)$^4_\Lambda\text{H}$ due to the earlier
$\Lambda$ freezeout.
From the detailed numbers listed in Table~\ref{LightNuclYield}, one can see
that in Au+Au collisions, the $dN/dy$ of $^4_\Lambda $H
($^4_{\overline{\Lambda}}\overline{\text{H}}$) is $1.48\times$10$^{-7}$
($5.12\times$10$^{-8}$) with FOAu-$\Lambda$ and $1.57\times$10$^{-6}$ ($5.43\times$10$^{-7}$)
with FOAu-$\Lambda^*$, implying the earlier $\Lambda$ freezeout enhances the
yields of both $^4_\Lambda $H and $^4_{\overline{\Lambda}}\overline{\text{H}}$ by a factor
of $10.6$. In Pb+Pb collisions,
the $dN/dy$ of $^4_\Lambda $H (same for $^4_{\overline{\Lambda}}\overline{\text{H}}$)
is $1.36\times$10$^{-7}$ with FOPb-$\Lambda$ and $5.43\times$10$^{-7}$ with FOPb-$\Lambda^*$, and
the enhancement factor due to the earlier $\Lambda$ freezeout is $4.0$.
It is interesting to see that
the $dN/dy$ of $^4_{\Lambda}$H ($^4_{\overline{\Lambda}}\overline{\text{H}}$)
at RHIC (with FOAu-$\Lambda^*$) is about $3.5 (1.0)$ times as large
as that at LHC (with FOPb-$\Lambda^*$), and the predicted yields
of the heavier $^4_{\Lambda}\text{H}$ ($^4_{\overline{\Lambda}} \overline{\text{H}}$) at RHIC
are larger than the measured $^4\text{He}$ ($^4\overline{\text{He}}$) yield at RHIC
(i.e., about $9.18\times10^{-7}$ for $^4\text{He}$ and
$2.91\times10^{-7}$ for $^4\overline{\text{He}}$~\cite{SunKJ15}).
Also the predicted $dN/dy$
of $^4_{\Lambda}\text{H}$ ($^4_{\overline{\Lambda}} \overline{\text{H}}$) with FOPb-$\Lambda^*$
at LHC (i.e., $5.43\times$10$^{-7}$) is very close to the measured $dN/dy$ of
$^4\text{He}$ ($^4\overline{\text{He}}$)
(i.e., about $7.8\pm3.1\times 10^{-7}$).
The larger yields of light (hyper)nuclei at lower colliding energies are also
observed in the predictions of thermal models (see, e.g., Ref.~\cite{And11}).
Compared with $^4\text{He}$ ($^4\overline{\text{He}}$), the larger or comparable
yields of $^4_{\Lambda}\text{H}$ ($^4_{\overline{\Lambda}} \overline{\text{H}}$)
are mainly due to the effects of earlier $\Lambda $ freezeout.
Generally, the yields of anti-(hyper)nuclei increase with the colliding energy,
and here that RHIC and LHC have the equal $dN/dy$ of $^4_{\overline{\Lambda}} \overline{\text{H}}$
is mainly due to the stronger earlier-$\Lambda $-freezeout effects at RHIC.
The future experimental measurement on $^4_{\Lambda}\text{H}$
($^4_{\overline{\Lambda}} \overline{\text{H}}$) would be very useful to test
the idea of earlier (anti-)$\Lambda$ freezeout.

\emph{5. Conclusion.}---%
The measured yield ratio $^3_{\Lambda}$H/$^3$He
($^3_{\overline{\Lambda}}\overline{\text{H}}$/$^3\overline{\text{He}}$)
in heavy-ion collisions at RHIC and LHC
can be naturally explained by the covariant coalescence model if the
(anti-)$\Lambda$ particles freeze out earlier than (anti-)nucleons but their
relative freezeout time is closer at LHC than at RHIC.
The earlier (anti-)$\Lambda$ freezeout
can significantly enhance the yield of $^4_{\Lambda}$H
($^4_{\overline{\Lambda}}\overline{\text{H}}$), leading to that
$^4_{\overline{\Lambda}}\overline{\text{H}}$ provides an easily measured candidate
for antimatter heavier than $^4\overline{\text{He}}$.
The larger relative p-$\Lambda$ ($\overline {\text p}$-$\overline {\Lambda}$)
freezeout time difference at RHIC leads to a larger (equal)
yield of $^4_{\Lambda}$H ($^4_{\overline{\Lambda}}\overline{\text{H}}$)
at RHIC than at LHC.
In future, more precise
measurement on $^3_{\Lambda}$H/$^3$He
and $^3_{\overline{\Lambda}}\overline{\text{H}}$/$^3\overline{\text{He}}$
as well as the measurement on $^4_{\Lambda}$H and $^4_{\overline{\Lambda}}\overline{\text{H}}$
would be extremely helpful to test the proposed freezeout scenario for (anti-)$\Lambda$
particles and the predictions on light (anti-)hypernuclei production presented in this work.

\emph{Acknowledgments.}---%
We are grateful to Vincenzo Greco, Che Ming Ko, Yu-Gang Ma, Zhang-Bu Xu,
Zhong-Bao Yin and Xian-Rong Zhou for helpful discussions.
This work was supported in part by the Major State Basic
Research Development Program (973 Program) in China under Contract Nos.
2015CB856904 and 2013CB834405, the NSFC under Grant Nos. 11275125
and 11135011, the ``Shu Guang" project supported by Shanghai Municipal
Education Commission and Shanghai Education Development
Foundation, the Program for Professor of Special Appointment (Eastern Scholar)
at Shanghai Institutions of Higher Learning, and the Science and Technology
Commission of Shanghai Municipality under Grant No. 11DZ2260700.

\end{document}